\documentclass[
 reprint,
superscriptaddress,
 amsmath,amssymb,
 aps,
pra,
longbibliography
]{revtex4-1}

\usepackage{siunitx}
\usepackage{graphicx}
\usepackage{epsfig}
\usepackage{float}
\usepackage{lipsum}
\usepackage[margin=2cm,includefoot]{geometry}
\usepackage[colorlinks, linkcolor=black]{hyperref}
\usepackage[english]{babel}
\usepackage{fancyhdr}
\usepackage{subfiles}
\usepackage{amsthm}
\usepackage{amsfonts}
\usepackage{braket}
\usepackage{endnotes}
\usepackage{physics}
\usepackage{blindtext}
\usepackage{multirow}
\usepackage{array}
\usepackage{tabularx}
\usepackage{cleveref}

\newcommand{\mycomment}[1]{}

\begin{document}

\title{Time resolution at the quantum limit of two incoherent sources based on frequency resolved two-photon-interference}

\author{Salvatore Muratore}
\affiliation{School of Mathematics and Physics, University of Portsmouth, Portsmouth PO1 3QL, UK}
\affiliation{Quantum Science and Technology Hub, University of Portsmouth, Portsmouth P01 3QL, UK}

\author{Vincenzo Tamma}
\email{vincenzo.tamma@port.ac.uk}
\affiliation{School of Mathematics and Physics, University of Portsmouth, Portsmouth PO1 3QL, UK}
\affiliation{Institute of Cosmology and Gravitation, University of Portsmouth, Portsmouth PO1 3FX, UK}
\affiliation{Quantum Science and Technology Hub, University of Portsmouth, Portsmouth P01 3QL, UK}

\date{\today}

\begin{abstract}
The Rayleigh criterion is a widely known limit in the resolution of incoherent sources with classical measurements in the spatial domain. Unsurprisingly the estimation of the time delay between two weak incoherent signals is afflicted by an analogue problem. In this work, we show the emergence of two-photon quantum beats in the frequency domain from the interference at a beam splitter of a photon emitted by a reference source and one from the two incoherent weak signals. We demonstrate, based on this phenomena, that with a relatively low number of measurements of the frequencies of the interfering photons either bunching or antibunching at the beam splitter output one can achieve a precision amounting to half of the quantum limit, independently of both the mode structure of the photonic wavepackets and the time delay to be estimated. The feasibility of the technique makes it applicable in astronomy, microscopy, remote clocks synchronization and radar ranging.
\end{abstract}

\maketitle

\paragraph*{Introduction}
The estimation of the time delay between two signals has a wide spectrum of applications, such as medical imaging~\cite{drexler2008optical}, distance estimation by time of flight measurements~\cite{hansard2012time}, gravitational waves detection~\cite{PhysRevLett.110.171102}.
However, in those scenarios in which the signals emitted from the sources could be purely incoherent commonly employed direct time resolved detection methods drastically fail, as they are indeed unable to precisely estimate the time delay between the two signals when it becomes much smaller than the temporal width of the optical field~\cite{BRADLEY1971391}, a situation analogous to the one in the spatial domain due to the Rayleigh's limit~\cite{BornWolf1999}. Nevertheless, thanks to the tools of quantum metrology~\cite{Cramer1999}, it has been shown that this is not a fundamental limit but instead a limitation of the current methods~\cite{PRXQuantum.2.010301}, and that the highest resolution achievable is the same at all scales, meaning that it is possible in theory to devise a high precision sensing scheme free from such limit. As a consequence, the research for a solution has led to the development of new approaches, among which the state-of-art is represented by temporal modes decomposition methods~\cite{PRXQuantum.2.010301, PhysRevA.105.062603, PhysRevResearch.3.033082}. 
This technique relies on a mode selective projection of the wavepacket of the incoming photons into orthogonal temporal modes, performed by mixing the incoming signals with specifically constructed gating pulses through a properly engineered waveguide, after which photon counting measurements are performed at the output. However, even if the technique promises to achieve high precision, it depends on the decomposition of the signal based on the specific temporal structure of the incoming photon wavepacket, which increases the difficulty in the experimental implementation~\cite{gosalia2025quantum}, and the presence of mode crosstalk affects the sensitivity~\cite{Tsang2016, Paur2016, Linowski2023, Santamaria2023, Santamaria2024}.
In the search of a new method which could enable the estimation problem in exam without the drawbacks of the current ones, we delve into the field of two-photon quantum interference, which has been a fruitful ground for many noteworthy results in quantum sensing~\cite{Hong1987, Shih1988, Lyons2018, Harnchaiwat2020, Sgobba2023}. In particular, in the celebrated Hong-Ou-Mandel quantum interference effect when two identical photons are sent to the two faces of a balanced beam splitter they will always bunch together at the same output port~\cite{Hong1987,Shih1988}. Instead, when the two photons are not identical, there is a non-zero probability that a coincidence event could happen.
Unfortunately, an estimation scheme solely based on this principle would require an overlap in time between the wavepackets of the two interfering photons, otherwise it would be insensitive to the parameter to estimate~\cite{Hong1987, Lyons2018}. However, new techniques based on multiphoton interference with inner variable boson sampling, which have first revealed an advantage in quantum computational complexity~\cite{ tamma2015multiboson, wang2018experimental, laibacher2018symmetries, tamma2014multiboson}, have shown that it is possible to surpass these inconveniences, while at the same time reaching a precision equal to the quantum Cram\'er-Rao bound in the estimation of different photonic parameters~\cite{Triggiani2023, Triggiani2024, maggio2025multi, legero2004quantum, legero2003time, PhysRevApplied.23.054033, 6xy6-c2yd, maggio2026ultimate, maggio2025quantum}. However, the exploration of novel two-photon interference techniques enabling  the resolution  in time of two incoherent signals beyond the Rayleigh limit in the time domain is yet largely terra incognita.\\
In this letter, we propose a new scheme for the estimation of the delay of two incoherent signals emitted by two weak thermal sources based on the quantum interference between a reference photon and one probe photon coming from either of the two weak sources, which overcomes the drawbacks of the aforementioned methods. We first demonstrate the unique quantum beat phenomena associated with the interference at a balanced beam splitter between the reference single photon and the probe photon in an incoherent superposition in the source emission times when they are collected by two cameras resolving their frequencies. We then exploit such interesting quantum phenomena to achieve  quantum sensitivity in the measurement of the unknown time delay. Our method does not require the mode decomposition of the wavepackets of the incoming photons and it works independently of their mode structure. Furthermore, we demonstrate that for indistinguishable photons at the detectors our scheme achieves constant precision at half of the quantum limit in the estimation of the time delay independently of the value of the delay. 
The feasibility of our technique makes it an ideal candidate in many applications such as remote clock synchronization~\cite{giorgetta2013optical, GAO2015773}, astronomical observations~\cite{krehlik2017fibre}, spontaneous emission time measurements~\cite{PhysRevA.105.062603}, condensed matter physics~\cite{RevModPhys.81.163}, or observation of incoherent scattering between photons in biological samples~\cite{chen2020extended}. It can also be applied in those scenarios where an incoherent superposition in time of the signals arises from the interaction of a coherent laser pulse with two displaced objects in presence of turbulence between the two objects or in the case in which their distance is higher than the coherence lenght of the two wavepackets. As an example, a new method has been proposed for the radar ranging problem~\cite{PhysRevApplied.20.064046, PhysRevResearch.6.033341, PhysRevLett.131.053803}, which works by sending a signal to the two objects of interest and measuring the returning photons, relying on their coherence. However, in this case the associated quantum Fisher information vanishes quadratically when $\Delta t\rightarrow0$, even in perfect experimental conditions. Instead, by employing our technique an incoherent scattering with the two objects would be advantageous instead of being a limitation, and constant Fisher information, beyond the classical limit, could be reached even in the case of small delays and independently of any turbulence which may contribute to generate incoherence \cite{abdukirim2023effects, cao2020compensation, xue2022inverse}.\\

\begin{figure}
\centering
\includegraphics[width=.9\columnwidth]{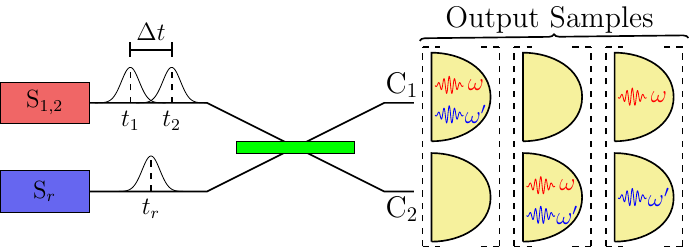}
\caption{Two-photon interferometer. Two incoherent photons are emitted with a time delay $\Delta t$ in the same line of sight, one of them interferes on a balanced beam splitter with a photon produced by a reference source. The possible outputs depend on the frequency of the detected photons and if they bunch together or hit different cameras. This scheme allows to retrieve the value of the time delay $\Delta t$ with quantum-enhanced sensitivity and a relatively low number of measurements.}
\label{fig:Setup}
\end{figure}

\paragraph*{Quantum interference of a single photon associated with with two weak incoherent delayed signals and a reference photon}

As shown in Fig.~\ref{fig:Setup} two incoherent weak signals propagate along the same line of sight separated by a time delay $\Delta t$. The associated quantum state in the single-photon regime, can be described by the density operator:
\begin{gather}
\begin{gathered}
\hat{\rho}=\frac{1}{2}\left(\ket{\psi_{1}}\bra{\psi_{1}}+\ket{\psi_{2}}\bra{\psi_{2}}\right),\\
\ket{\psi_{i}}=\int_{\mathbb{R}}\dd \omega\ \xi_{i}(\omega)\hat{a}_{1}^{\dagger}(\omega)\ket{\mathrm{vac}}, \quad i=1,2
\end{gathered},
\label{eq:SourcePhoton}
\end{gather}
with $\xi_{i}(\omega)=\overline{\xi}(\omega)\mathrm{e}^{-\mathrm{i}\omega t_{i}},\, i=1,2$ the frequency probability amplitude of the photons, $t_{i}$ the central time at which the $i$th photon hits the beam splitter, with the time delay $\Delta t=t_{2}-t_{1}$ and the average time $t_{s}=(t_{1}+t_{2})/2$, while $\hat{a}_1^{\dagger}(\omega)$ is the bosonic creation operator associated with a photon with frequency $\omega$.
We notice that $\ket{\psi_1}\, \text{and}\, \ket{\psi_2}$ are not orthogonal.\\
The probe photon emitted by the two incoherent sources impinges on one of the two faces of a balanced beam splitter, while a reference photon is injected at the other input of the beam splitter at central time $t_{r}$ with state
\begin{equation}
\ket{\psi_{0}}=\int_{\mathbb{R}}\dd \omega\ \xi_{0}(\omega)\hat{d}_{0}^{\dagger}(\omega)\ket{\mathrm{vac}}, 
\label{eq:RefPhoton}
\end{equation}
where $\hat{d}_0^\dag(\omega)$ is the bosonic creation operator associated with a reference photon at the frequency $\omega$, and $\xi_{0}=\overline{\xi}\mathrm{e}^{-\mathrm{i}\omega t_{r}}$.\\
The bosonic operators satisfy the commutation relations $\left[\hat{a}_{S}(\omega), \hat{d}^{\dagger}_{S'}(\omega')\right] = \sqrt{\nu}\,\delta_{S,S'}\delta(\omega-\omega')$, with $S,S' = 0,1$, which means that we can rewrite $\hat{d}_{S}(\omega)=\sqrt{\nu}\hat{a}_{S}(\omega)+\sqrt{1-\nu}\hat{b}_{S}(\omega)$ with $\hat{a}$ and $\hat{b}$ orthogonal modes. Here, $0 \leq \nu \leq 1$ represents the degree of indistinguishability between photons in all physical degrees of freedom except for their arrival times, where $\nu = 1$ corresponds to complete indistinguishability. In all instances, the probability amplitudes in the frequency domain are such that $\left|\xi_{0}\right| = \left|\xi_{1}\right| = \left|\xi_{2}\right| = \overline{\xi}$.\\
After the interaction at the beam splitter, the detection is performed using two single-photon cameras positioned at the two output ports, allowing us to detect the frequencies $\omega, \omega'$ of the two photons.
A single measurement result, $(\Delta \omega, X)$, is defined by the difference in the frequency, $\Delta \omega=\omega-\omega'$, of the two photons and whether both photons impinge on the same camera ($X = B$, bunching event) or on different cameras ($X = A$, antibunching event).\\
\begin{figure}
\includegraphics[width=.75\columnwidth]{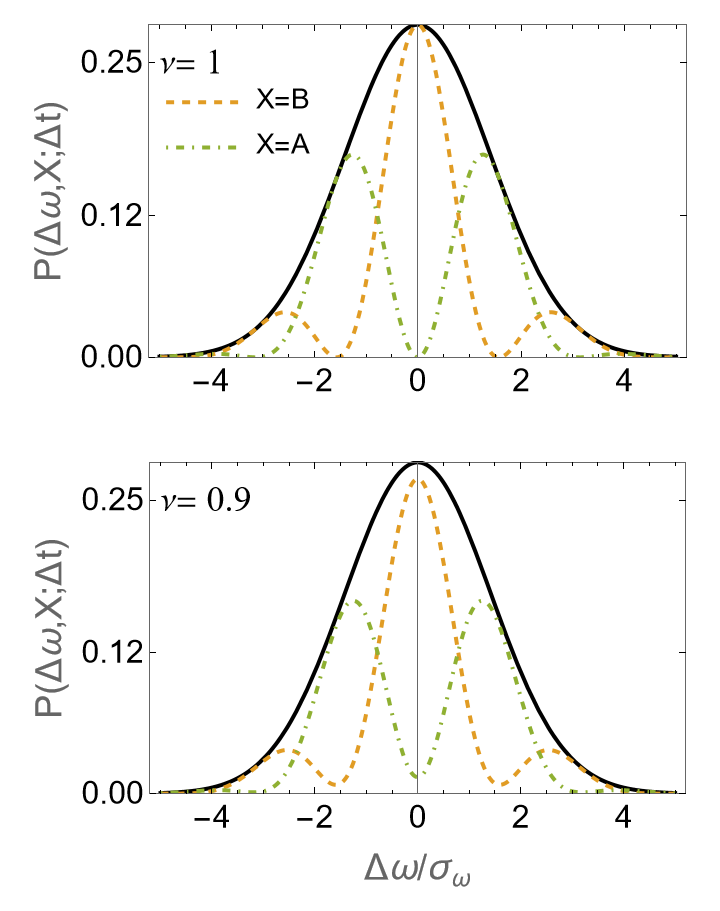}
\caption{Quantum beat interference in the probability distribution $P(\Delta\omega, X; \Delta t)$ at the output of the two-photon interferometer in Fig.~\ref{fig:Setup}. The probability given by Eq.~\eqref{Eq:deltafixed} is plotted as a function of the frequency difference $\Delta\omega$, in units of the frequency bandwidth $\sigma_{\omega}$, for a fixed time delay $\Delta t = 4/\sigma_{\omega}=8\sigma_{t}$ in the cases of indistinguishable photons, $\nu = 1$, and partially distinguishable photons, $\nu = 0.9$. The coincidence probability ($X = A$) in the dashed-dotted green line, and the bunching probability ($X = B$) in the dashed orange line. The probability clearly manifests quantum beats with a periodicity inversely proportional to $\Delta t$.
}
\label{fig:Probs}
\end{figure}
We derive the probability
\begin{multline}
P_{\nu}(\Delta \omega, X; \Delta t, t_{r}-t_{s}) = \frac{1}{2}\eta C(\Delta \omega)\\
\times\left\{1 + \alpha(X)\nu\cos\left[\Delta \omega\frac{\Delta t}{2}\right]\cos\left[\Delta \omega(t_{r}-t_{s})\right]\right\},
\label{eq:Prob}
\end{multline}
where $\alpha(B) = 1$, $\alpha(A) = -1$, and $\eta < 1$ is a factor that accounts for detector losses, and $C(\Delta \omega)$ is an envelope function which shape is given by the frequency distribution, e.g., $C(\Delta\omega)= \exp\left(-\Delta \omega^{2}/4\sigma^{2}_{\omega}\right)/\sqrt{4\pi\sigma^{2}_{\omega}}$ in the case of a Gaussian frequency distribution $\xi(\omega)$ with variance $\sigma^{2}_{\omega}=1/4\sigma^{2}_{t}$, with $\sigma^{2}_{t}$ the variance of the corresponding photon temporal wavepacket \cite{Suppl}. This probability describes the observation of a photon pair in the sampling outcome $(\Delta \omega, X)$ and depends on the time delay $\Delta t = t_{1} - t_{2}$ and the time centroid $t_{s} = (t_{1} + t_{2}) / 2$ associated with the two incoherent signals. Importantly, this result indicates that two-photon interference beats can be observed as a function of the frequency difference $\Delta \omega$.\\
If the two incoherent pulses do not experience any time delay, i.e. they are emitted at the same time, $\Delta t=0$, the two-photon interference beatings in the probability in Eq.~\eqref{eq:Prob} are function of $\Delta \omega$ with a periodicity inversely proportional to the difference $(t_{r}-t_{s})$ between the central time of emission $t_{r}$ of the reference photon wave packet and the time centroid $t_{s}$ of the probe signals. Here we will consider the case of signals emitted at different times, $\Delta t\neq 0$, and a reference photon with central arrival time at the beam splitter input given by the centroid time of the probe photon, which can be estimated through time resolved direct detection~\cite{gosalia2025quantum}, so that $t_{r}=t_{s}$. In this regime, the quantum beats in the probability
\begin{equation}
\begin{aligned}
P_{\nu}(\Delta \omega, X; \Delta t) =&\,\frac{1}{2} \eta\,C(\Delta \omega)\\
&\times\left\{1 + \alpha(X)\nu\cos\left[\Delta \omega\frac{\Delta t}{2}\right]\right\},
\label{Eq:deltafixed}
\end{aligned}
\end{equation}
obtained from Eq.~\eqref{eq:Prob} under the condition $t_{r} = t_{s}$, exhibit a periodicity inversely proportional to the time delay $\Delta t$, as can be seen in Fig.~\ref{fig:Probs}. We aim to exploit this quantum feature of the scheme to estimate the time delay between the incoherent signals. Furthermore, since our measurement is carried out by resolving in the frequency, i.e. the conjugate variable of time, it is only necessary for the cameras to be able to resolve oscillations with a period $\propto 1/\Delta t$, bypassing the requirement of high precision in time, which is necessary in time-resolved direct detection methods. 

\paragraph*{Quantum estimation technique for indistinguishable photons}
We will now show that we can develop an efficient quantum metrological technique, harnessing two-photon quantum beats in the frequency domain, to estimate the time delay $\Delta t$, through sampling from the probability distribution given in Eq.~\eqref{Eq:deltafixed}.
The estimation of $\Delta t$ is done by recording the sampling outcomes $\{\Delta \omega, X\} = (\Delta \omega_{i}, X_{i})$, with $i = 1, \ldots, N$, obtained from $N$ two-photon interference measurements, using the interferometer illustrated in Fig.~\ref{fig:Setup}, and using the gathered data to evaluate the maximum-likelihood estimator $\widetilde{\Delta t}$.\\
Remarkably, we now show that with our technique it is possible to accurately estimate the time delay, $\Delta t$, between the two photons even when it approaches zero.
To this end, we evaluate the Cram\'er-Rao bound~\cite{Cramer1999}
\begin{equation}
\mathrm{Var}[\widetilde{\Delta t}] \geq \frac{1}{N F_{\nu}(\Delta t)},
\label{eq:CRB}
\end{equation}
which establishes the fundamental lower bound on the variance $\mathrm{Var}[\widetilde{\Delta t}]$ achievable using our metrological technique in a maximum likelihood estimation, with $N$ representing the number of detected photon pairs, while $F_{\nu}(\Delta t)$ is the Fisher information corresponding to the frequency-resolved measurement scheme.

\begin{figure*}[t]
  \centering
  \includegraphics[width=0.49\textwidth]{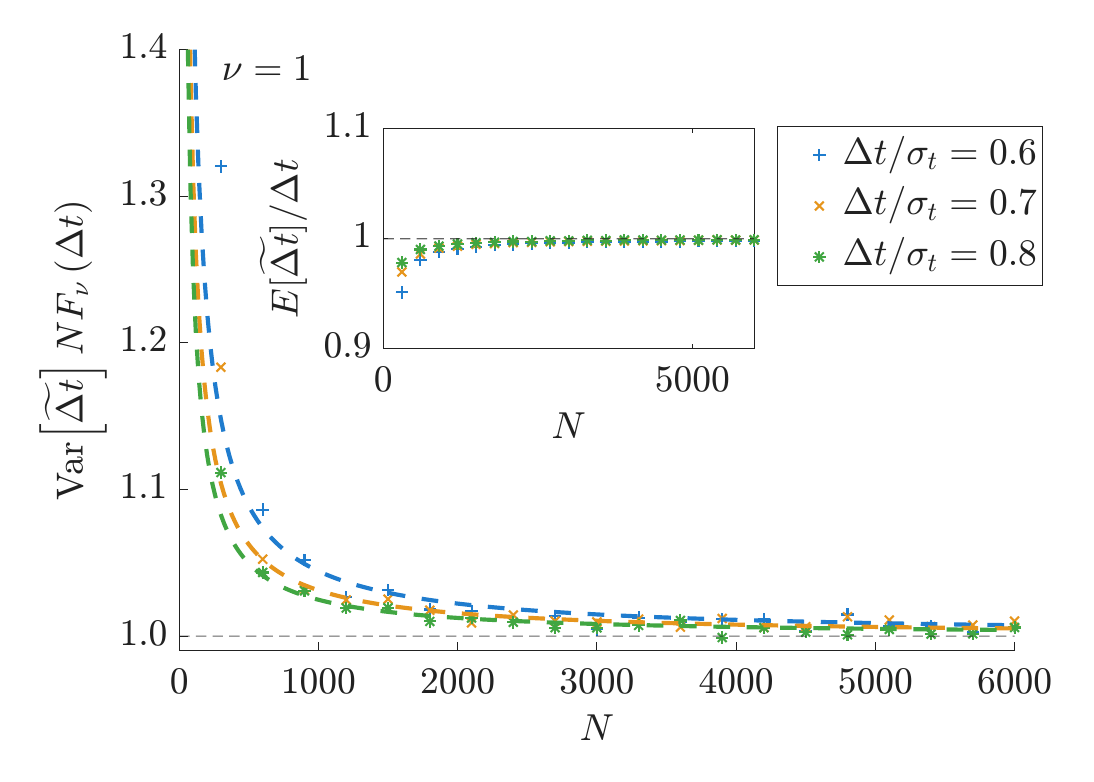}
  \includegraphics[width=0.49\textwidth]{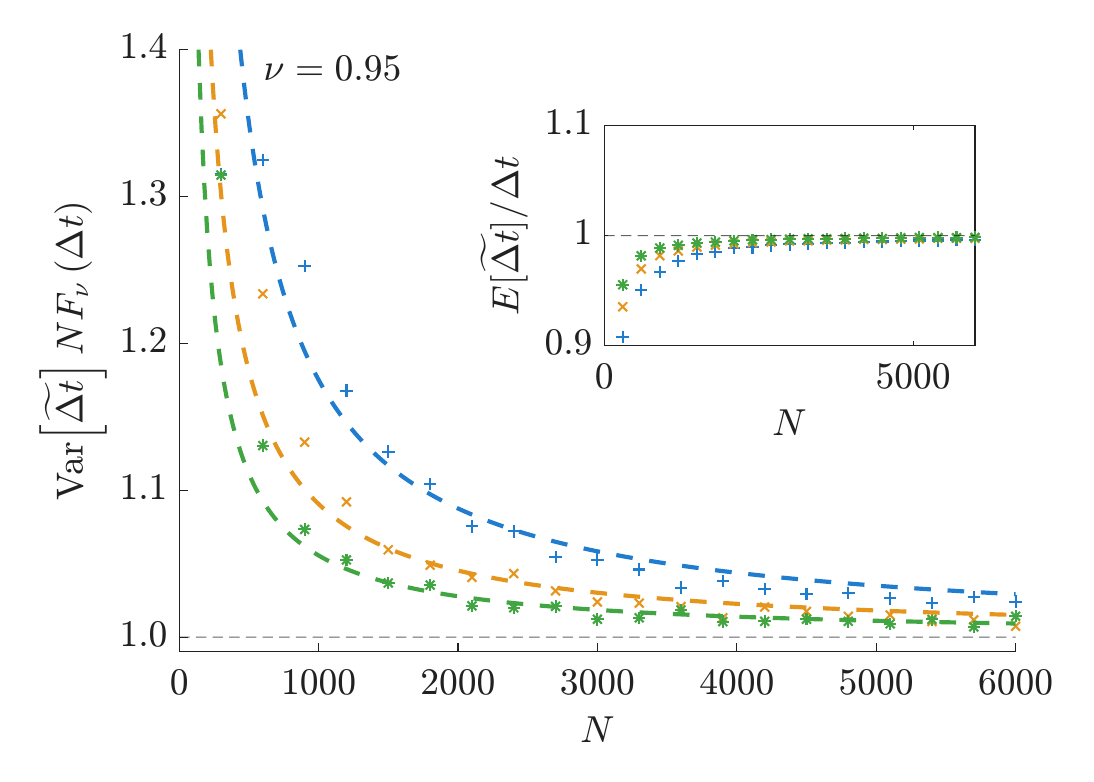}
\caption{Numerical simulations for Gaussian wavepackets of the rate of convergence of the variance of the maximum likelihood estimator $\widetilde{\Delta t}$ normalized to the Cram\'er-Rao bound in Eq.~\eqref{eq:CRB}, for different numbers $N$ of collected samples and for $\Delta t/\sigma_t=0.6,0.7,0.8$, in the case of indistinguishable photons, $\nu=1$ (Fisher information $F_{\nu=1}$ in Eq.~\eqref{eq:Fisher}), and partially distinguishable photons, $\nu=0.95$ (Fisher information $F_{\nu}(\Delta t)$ in Eq.~\eqref{eq:Fishres}). We show that the data can be fitted with the curve $1+a/N$, where $a/N$ is a correction term of order $1/N$ of the variance of the estimator normalized to the Cram\'er-Rao bound in Eq.~\eqref{eq:CRB} for $N\gg 1$. This shows that our scheme approaches the Cram\'er-Rao bound already for $N\simeq 5000$. In the insets, we also plot the estimated expectation value of the maximum-likelihood estimator normalized by its real value, showing that the estimation is unbiased already again for $N\simeq 5000$.}
\label{fig:Numeric}
\end{figure*}
If we were to use time-resolved direct measurements for the estimation of $\Delta t$ the Fisher information would vanish quadratically for $\Delta t\rightarrow 0$ even in ideal conditions such as unlimited time resolution, making it impossible to accurately estimate the parameter.
Contrarily we now demonstrate that the Fisher Information for our technique is constant for every value of $\Delta t$, even when it approaches zero.
We evaluate the Fisher information found in Eq.~\eqref{eq:CRB} for the probability distribution in Eq.~\eqref{Eq:deltafixed},  in the case of perfectly indistinguishable photons $\nu=1$, obtaining
\begin{equation}
F_{\nu=1}(\Delta t)=\eta\frac{\sigma^{2}_{\omega}}{2}=\eta\frac{1}{8\sigma^{2}_{t}},
\label{eq:Fisher}
\end{equation}
which is directly proportional to the variance $\sigma^{2}_{\omega}$ of the frequency distribution and inversely proportional to the variance $\sigma^{2}_{t}$ of the temporal wavepacket. Consequently, narrower temporal distributions lead to a higher achievable precision in the estimation of the time delay \cite{Suppl}. Notably, the independence of $F(\Delta t)$ from $\Delta t$ indicates that the estimation precision attainable with our two-photon interference scheme remains constant for all parameter values, irrespective of the photonic wavepacket structure. Furthermore, the Fisher Information differs from the quantum limit 
\begin{equation}
Q(\Delta t)= \sigma_{\omega}^{2}=\frac{1}{4\sigma_{t}^{2}}
\label{eq:Qufi}
\end{equation} 
only by the constant factor $\frac{1}{2}\eta$~\cite{PRXQuantum.2.010301}. To illustrate the significance of this result, consider a biphoton rate of $1\,\mathrm{MHz}$ with lossless detectors ($\eta=1$). If photons with a temporal width $\sigma_{t}\sim 60\,\mathrm{fs}$ are employed, which is achievable with current technology \cite{mosley2008heralded, nasr2008ultrabroadband}, a timing precision of the order of attoseconds could be reached in only $4\,\mathrm{h}$ of measurement, and for a $\sigma_{t}\simeq 10\,\mathrm{fs}$, it would only take about $8$ min to reach the same precision, making our method not only higly precise, but also remarkably fast. Furthermore our sampling scheme relyes on a simple state-of-the-art optical apparatus and is independent of the specific mode structure of the photonic wavepackets. It is therefore more experimentally feasible than other recent imaging techniques relying on demultiplexing, i.e. the decomposition of the incoming optical field into a basis of orthogonal modes dependent on the structure of the input wavepackets through the use of a specific waveguide, also wavepacket dependent, which by consequence renders the measurements affected by mode cross-talking ~\cite{Tsang2016, Paur2016, Santamaria2023, Santamaria2024}.
Furthermore, for photons with highly overlapping wavepackets, $\sigma_{\omega}\Delta t \ll 1$, we demonstrate that the same Fisher information given in Eq.~\eqref{eq:Fisher} can be also attained without frequency-resolved measurements by relying solely on the sampling outcomes ($X = A,B$) obtained with two simple bucket detectors \cite{Suppl}. We estimate the parameter $\Delta t$ by using standard maximum likelihood estimation. In particular, in Fig.~\ref{fig:Numeric} (a) are depicted the numerical simulations of the variance, $\text{var}(\Delta \tilde{t})$, of the maximum likelihood estimator $\Delta \tilde{t}$ for a Gaussian wavepacket normalized by the Cram\'er-Rao bound in Eq.~\eqref{eq:CRB}, and the expected value of the estimator, $E[\Delta \tilde{t}]$, normalized to the real value of the parameter $\Delta t$ \cite{Suppl}. The inset confirms the unbiased character of the estimator, while the main graph displays a fit of the numerical data to the function $1 + a/N$, where $a/N$ is the correction term of order $1/N$ of the variance normalized to the Cram\'er-Rao bound, as shown in the supplemental material. The parameters reported in Table~\ref{table:1} reveal that higher-order terms in $1/N$ are negligible. For $\nu=1$ our method approaches the Cram\'{e}r-Rao bound in Eq.~\eqref{eq:CRB}, apart from a correction term $a/N$ of order $\simeq 1\%$, already with $N \simeq 5000 $ detected photon pairs, as evident in Fig.\ref{fig:Numeric} (a).\\

\begin{table}[h!]
\centering
\resizebox{0.85\columnwidth}{!}{%
\renewcommand{\arraystretch}{2}
\begin{tabular}{ |c|c|c|c|}
\hline
\multicolumn{4}{|c|}{$\nu=1$}\\
\hline
 $\Delta t/\sigma_{t}$ & SSE & $R^{2}$ & $a$ $(a_{min},a_{max})$ \\ 
 \hline
 \,\,\,\,0.6 \,\,\,\,& \,\,\,\, 6.2531e-04 \,\,\,\,& \,\,\,\,0.9150 \,\,\,\,& \,\,\,\,44.34\, (39.52, 49.15)\,\,\,\,\\  
 \hline
 0.7 & 1.9697e-04 & 0.9178 & 31.26\,  (28.56, 33.96) \\
 \hline
 0.8 & 1.7529e-04 & 0.9216 & 24.79\, (22.24, 27.34)\\
 \hline
\multicolumn{4}{|c|}{$\nu=0.95$}\\
\hline
 $\Delta t/\sigma_{t}$ & SSE & $R^{2}$ & $a$ $(a_{min},a_{max})$ \\ 
 \hline
 0.6 & 0.0015 & 0.9429 & 175.2 \, (162.3, 188.1) \\  
 \hline
 0.7 & 6.5066e-04 & 0.9161 & 90.64 \, (82.28, 99.01) \\
 \hline
 0.8 & 2.2080e-04 & 0.9070 & 55.6 \, (50.73, 60.47)\\
 \hline
\end{tabular}
}
\caption{In this table we show the goodness of the fits plotted in Fig.~\ref{fig:Numeric}, for indistinguishable photons ($\nu=1$) and for partially distinguishable photons ($\nu=0.95$). In the first column, we insert the values of the parameter to estimate in units of $\sigma_{t}$. For each of these values, we associate the summed square of residuals (SSE), the R-square ($R^{2}$), and the coefficient $a$, evaluated with $95\%$ confidence $(a_{min}, a_{max})$.}
\label{table:1}
\end{table}

\paragraph*{Quantum estimation technique for partially distinguishable photons}

\begin{figure}
\includegraphics[width=1\columnwidth]{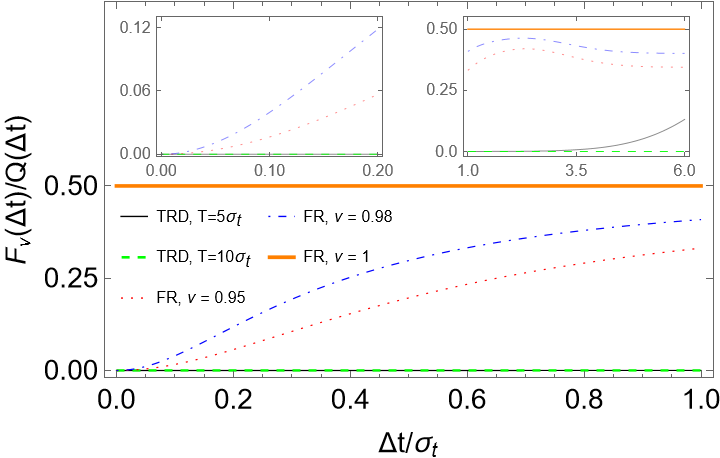}
\caption{Plots of the Fisher information $F_{\nu}(\Delta  t)$ in Eq.~\eqref{eq:Fishres} for frequency-resolved (FR) measurements for $\eta=1$, and Fisher information for time-resolved direct (TRD) detection as a function of the time delay $\Delta t/\sigma_{t}$, normalized by the Quantum Fisher information in Eq.~\eqref{eq:Qufi}, considering as an example a Gaussian temporal wavepacket with unitary $\sigma_{t}$. Direct detection in solid black line and dashed green line for a direct detection with resolution $T=5\sigma_{t}$ and $T=10\sigma_{t}$ respectively, $F_{\nu}$ in Eq.~\ref{eq:Fishres} in dotted red, $\nu=0.95$, dot-dashed blue, $\nu=0.98$, and thick solid orange, $\nu=1$. In the left inset we show the same curves (constant function $F_{\nu=1}/Q=1/2$ omitted here) for small delays,$\Delta t/\sigma_{t}<0.2$, while in the right inset we show them for large delays, $1<\Delta t/\sigma_{t}<6$, highlighting the clear advantage of our method in both regimes. We observe that for both the values $\nu=0.95,0.98$ the Fisher information $F_{\nu}(\Delta t)$ approaches its maximum for $\Delta t/\sigma_{t}\simeq 2$ and maintains a constant value given by Eq.~\eqref{eq:Fishlimit} already for $\Delta t/\sigma_{t}\simeq 5$.
}
\label{fig:comparison}
\end{figure}

In experimental implementations, photons impinging on the detectors may exhibit partial distinguishability apart from the overlap in their temporal wavepacket, $\nu < 1$. Nonetheless, we can prove that the quantum advantage offered by our technique over time-resolved direct detection still remains. We consider for simplicity the case of Gaussian wavepackets with unitary $\sigma_t$.\\
The Fisher information for our method, for $\nu<1$, can be expressed as \cite{Suppl}
\begin{equation} 
\begin{aligned} 
F_{\nu}(\Delta t)=\frac{1}{4}\eta&\int\dd\Delta \omega C(\Delta \omega)(\Delta \omega)^2\\ &\times\frac{\nu^{2}\sin^{2}\left[\Delta \omega\Delta t/2\right]}{1-\nu^{2}\cos^{2}\left[\Delta \omega\Delta t/2 \right]}, \end{aligned} 
\label{eq:Fishres}
\end{equation}
which, in the regime of large delays $\Delta t\gg\sigma_{t}$, approaches the constant value
\begin{equation}
F_{\nu}(\Delta t\gg\sigma_{t})=\left(1-\sqrt{1-\nu^{2}}\right)F_{\nu=1},
\label{eq:Fishlimit}
\end{equation}
with $F_{\nu=1}$ in Eq.~\eqref{eq:Fisher}~\cite{Suppl}. Remarkably, this shows that the achievable precision in this regime is again constant and independent of the estimated parameter $\Delta t$, and increases for narrower temporal distributions.
To prove the advantage of our method in Fig.~\ref{fig:comparison} we compare the Fisher information in Eq.\eqref{eq:Fishres} with a realistic time-resolved direct-detection scenario in which the temporal resolution $T$ is finite. 
We examine two different cases associated with temporal resolutions $T=5\sigma_{t}$ and $T=10\sigma_{t}$. If we consider, for example, applications related to photon scattering in biological tissue, where the detected field temporal width is in the picoseconds range, and to thermal black-body radiation, whose coherence time is typically tens to hundreds of femtoseconds \cite{kholiqov2020time, ricketti2022coherence}, this would require detectors with quite high resolution $T\approx 10^{1}\text{--}10^{2}\,\mathrm{ps}$ \cite{korzh2020demonstration, gao2015measurement}. Furthermore, as shown in Fig.~\ref{fig:comparison}, in both cases the Fisher information obtained from direct detection would be smaller than the one achieved by our method, and already for $T=10\sigma_{t}$ the standard direct detection approach becomes effectively unusable for any value of $\Delta t$ given the vanishing Fisher information.
Again, we show in Fig.~\ref{fig:Numeric} (b) that the estimator is unbiased and the convergence of the variance of the estimator to the Cram\'er-Rao bound apart from a correction term of the order $1/N$.\\
\paragraph*{Conclusions}

The interferometric scheme we introduced in this work has the peculiar trait of manifesting the two-photon quantum beat interference phenomenon, which emerges from the frequency-resolved measurements of a single reference photon interfering with another photon from two incoherent weak signals. 
Moreover, we showed that this purely quantum feature is the basis upon which we can build a method to achieve for indistinguishable photons constant precision in the estimation of the time delay between the two incoherent signals, including the small delay regime in which classical time-resolved direct detection methods fail.
Remarkably, this is possible by performing simple frequency-resolved sampling measurements from the output probability distribution of the photons. We also highlight that the achieved precision differs from the quantum limit only by a constant factor even when imperfect detectors are employed, and we can saturate the CRB already with a relatively low number of measurements.
The advantage of the proposed scheme lies in its remarkable simplicity, which makes it possible to obtain constant precision without the complexities of the decomposition in orthogonal modes and the associated cross-talk drawbacks, maintaining the same efficiency independently of the value of the delay to estimate and of the mode structure of the photonic wavepackets.
Furthermore, we also showed that even for partially distinguishable photons our technique can outperform classical detection, including the small delay regime.
For its experimental feasibility this technique lends itself to important applications in astronomy, remote clock synchronization and radar ranging.

\section*{ACKNOWLEDGMENTS}

We thank Paolo Facchi and Luca Maggio for the helpful discussions. This project was partially supported by the Air Force Office
of Scientific Research under award number FA8655-23-1-
7046.

\bibliography{bibliography}

\onecolumngrid

\end{document}